# THE 2011 OCTOBER DRACONIDS OUTBURST. II. METEOROID CHEMICAL ABUNDANCES FROM FIREBALL SPECTROSCOPY

José M. Madiedo[1, 2], Josep M. Trigo-Rodríguez[3], Natalia Konovalova[4], Iwan P. Williams[5], Alberto J. Castro-Tirado[6], José L. Ortiz[6] and Jesús Cabrera-Caño[2]

[1] Facultad de Ciencias Experimentales, Universidad de Huelva, 21071 Huelva, Spain.
[2] Departamento de Física Atómica, Molecular y Nuclear. Facultad de Física. Universidad de Sevilla. 41012 Sevilla, Spain.
[3] Institute of Space Sciences (CSIC-IEEC). Campus UAB, Facultat de Ciències, Torre C5-p2. 08193 Bellaterra, Spain.
[4] Institute of Astrophysics of the Academy of Sciences of the Republic of Tajikistan, Bukhoro, str. 22, Dushanbe 734042, Tajikistan.
[5] School of Physics and Astronomy, Queen Mary, University of London, London, UK.
[6] Instituto de Astrofísica de Andalucía, CSIC, Apt. 3004, 18080 Granada, Spain.

**ABSTRACT**
On October 8, 2011 the Earth crossed dust trails ejected from comet 21P/Giacobini-Zinner in the late 19th and early 20$^{th}$ Century. This gave rise to an outburst in the activity of the October Draconid meteor shower, and an international team was organized to analyze this event. The SPanish Meteor Network (SPMN) joined this initiative and recorded the October Draconids by means of low light level CCD cameras. In addition, spectroscopic observations were carried out. Tens of multi-station meteor trails were recorded, including an extraordinarily bright October Draconid fireball (absolute mag. -10.5) that was simultaneously imaged from three SPMN meteor observing stations located in Andalusia. Its spectrum was obtained, showing a clear evolution in the relative intensity of emission lines as the fireball penetrated deeper into the atmosphere. Here we focus on the analysis of this remarkable spectrum, but also discuss the atmospheric trajectory, atmospheric penetration, and orbital data computed for this bolide which was probably released during 21P/Giacobini-Zinner return to perihelion in 1907. The spectrum is discussed together with the tensile strength for the October Draconid meteoroids. The chemical profile evolution of the main rocky elements for this extremely bright bolide is compared with the elemental abundances obtained for 5 October Draconid fireballs also recorded during our spectroscopic campaign but observed only at a single station. Significant chemical heterogeneity between the small meteoroids is found as we should expect for cometary aggregates being formed by diverse dust components.

# 1 INTRODUCTION
The parent body of the October Draconid meteoroid stream is the short period comet 21P/Giacobini-Zinner (hereafter 21P). This young meteoroid stream produces an annual display of meteors from about October 6 to October 10, with a maximum activity around October 8 (Jenniskens 2006). Most years the October Draconids can be catalogued as a minor meteor shower, but from time to time it produces brief but spectacular meteor storms. Impressive October Draconid displays have been observed in the last centuries, but among the strongest ones were those that occurred in 1933 and 1946 (Comas Solà 1939; Jacchia et al. 1959).

As a comet approaches the Sun, the cometary surfaces heats up, producing high sublimation rates that cause significant outgassing, and a massive release of particles in the micron to millimeter size range (Whipple, 1951). Modern computational techniques are able to model the dynamic evolution of dust particles released by comets during perihelion approaches to the Sun. Using such methods, Wu & Williams (1995) showed that in the case of the October Draconids, the dust had to be released close to the present time so that the stream is very young. Recently, several researchers predicted that on October 8, 2011 the Earth would encounter dust trails ejected by comet 21P in the late 19th and early 20th centuries. An outburst with an activity of several hundred meteors per hour was forecasted by Vaubaillon et al. (2010) and Maslov et al. (2011). The SPanish Meteor Network (SPMN) joined an international October Draconid monitoring initiative to observe this event. Some of the meteor observing stations operated by the SPMN set up additional high-sensitivity CCD video cameras in order to provide a better coverage of meteor activity. The October Draconid meteoroids are known to be very fragile (Trigo-Rodríguez and Llorca 2006; Maslov et al. 2011), and obtaining accurate data is essential in order to gain a better understanding of the physico-chemical properties of these particles. Holographic diffraction gratings were attached to some of our cameras in order to obtain the emission spectra of the October Draconids so that the column temperature, electron density, and chemical compositional data as a function of height could be inferred. Unfortunately, the Moon was at a phase of about 91%, so that it interfered with the observation. Despite this, meteors as faint as mag. +1/+2 were recorded together with some fireballs at multiple stations.

Spectra of ten October Draconids with magnitudes ranging from 0 to -3 have been previously analyzed by Millman (1972), who suggested that the meteoroids had a relative distribution by weight of sodium, calcium, iron and magnesium very similar to the carbonaceous chondrites and to the olivine-bronzite chondrites. More recently, from the analysis of a mag. -4 photographic fireball, Borovicka et al. (2007) concluded that the relative abundances of Na, Mg, and Fe in these meteoroids were nearly chondritic, with differential ablation causing preferential loss of sodium at the beginning of the atmospheric trajectory. Here we present analysis of an extraordinarily bright October Draconid event (absolute magnitude -10.5) recorded at three SPMN stations. The emission spectrum of this unusual bolide is unique as, to our knowledge, an October Draconid fireball of this brightness has never been recorded with a spectrograph. This fireball is called SPMN081011_194759. The computed trajectory and velocity when compared with the relative abundances of the main rocky elements provides unprecedented information on differential ablation along the meteor path (Trigo-Rodríguez et al. 2003) and clearly supports the idea that ablation of different chemical elements occurs at different locations along the meteor column. The spectrum of this fireball is compared with spectra produced by five fainter single-station October Draconid fireballs which were also imaged during the maximum peak of the 2011 outburst. Due to the fragile nature of the October Draconid meteoroids, most of these faint spectra are of the brightest flares typically associated with meteoroid disintegration.

## 2 INSTRUMENTATION AND DATA REDUCTION TECHNIQUES

Four of the SPMN stations located in Andalusia -Sevilla, Cerro Negro, El Arenosillo and Sierra Nevada, whose locations are listed in Table 1- were involved in the observations of the October Draconid fireballs considered here. These use unintensified low-light monochrome CCD video systems to analyze meteor activity (Madiedo & Trigo-Rodríguez 2007; Trigo-Rodríguez et al. 2007). These CCD video cameras generate interlaced video imagery at 25 frames per second with a resolution of 720x576 pixels (PAL video system). Traditionally, these SPMN video meteor stations employed between 4 and 9 high-sensitivity Watec CCD video cameras (models 902H and 902H Ultimate from Watec Corporation, Japan) to monitor the night sky. However, during this campaign an additional effort was made in order to increase the number of imaging devices at some of these stations. The equipment operating from every station is detailed in Table 1.

The video cameras are equipped with a 1/2" Sony interline transfer CCD image sensor with their minimum lux rating ranging from 0.01 to 0.0001 lux at f1.4. Aspherical fast lenses with focal lengths ranging from 4 to 25 mm and focal ratios between 1.2 and 0.8 are used for the imaging objective lens. In this way, different areas of the sky can be covered by every camera and point-like star images are obtained across the entire field of view. This configuration allows us to record meteors brighter than +3±1 stellar magnitude under dark skies. The cameras are connected to a computer via a video acquisition card. Different PCI and USB 2.0 video acquisition cards have also been tested and employed. In most cases, we found a better performance using internal PCI cards, USB acquisition cards tend to fail or get damaged more often, although they are very useful mainly for mobile stations based on portable computers. The computers use the UFOCapture software (Sonotaco, Japan) to automatically detect meteor trails and store the corresponding video sequences on hard disk. On the other hand, the cameras are arranged in such a way that the common atmospheric area monitored by neighbouring stations is maximized. A more detailed description of these systems has been given elsewhere (Madiedo & Trigo-Rodríguez 2010; Madiedo et al. 2010).

Since 2006 one of the stations (Cerro Negro) has been set-up as a mobile system operating when necessary in a dark countryside environment at about 60 km north from Seville. For this October Draconid outburst, low-scan CCD cameras and CCD video devices were installed there (Table 1). Power consumption is a major drawback for the operation of this station, and this is solved by using low consumption computers (netbooks and PC desktops to minimize power requirements) fed by an array of 12-volts car batteries. Fortunately during the whole October Draconid observing campaign the imaging devices operated far enough from the dew point, so that the dew removal systems we employ to prevent humidity condensation on the optics, which imply a significant increase in power consumption, were not necessary. In this way, with the aid of 8 lead batteries with a capacity of 74Ah , this mobile station worked continuously, from dusk on Oct. 8 until sunrise on Oct. 9. Five CCD video cameras operated from Cerro Negro with attached transmission holographic diffraction gratings (500 or 1000 lines/mm, depending on the device) to obtain the emission spectra resulting from the ablation of meteoroids in the atmosphere. These spectra provide chemical information about these particles of interplanetary matter (Borovička 1993; Trigo-Rodríguez et al. 2003, 2004, 2009). Due to the quality of the night sky in this location, this station is usually setup when SPMN spectroscopic campaigns are organized. For the analysis of the 2011 October Draconid outburst, two high-resolution spectrographs were also installed in Cerro Negro. Unfortunately because of the lunar phase (91%), the images taken by these devices got saturated immediately and so they were disconnected to save energy.

Stations #2, 3 and 4 work in an autonomous way by means of software developed by us. They are automatically switched on and off at sunset and sunrise, respectively. During the observing session a software package tests that the video capture system is working, and restarts it in the event of failure. This software also removes images corresponding to false detections, such as cosmic rays. It also estimates the brightness of every recorded event, and when a very bright fireball is imaged, an email is automatically sent to the operator. Finally, at the close of the observing session, data recorded during the night are automatically compressed and sent to our FTP server. This is not the case for the mobile station, where the video files are manually saved to the server's hard disk. Once the images recorded by every station are stored on the FTP server, another software package identifies trails that were simultaneously recorded from at least two different stations. A copy of these multi-station data is placed in a separate folder where the video frames on every video file are co-added in order to increase the number of stars available for the astrometric measurement. A composite image showing the whole meteor trail is also generated for every event.

Once multi-station events have been identified and pre-processed as described above, an astrometric measurement is carried out by hand in order to obtain the coordinates of the meteor along its apparent path from each station. The AMALTHEA software (Trigo-Rodríguez et al. 2009; Madiedo et al. 2011), transforms these coordinates into equatorial coordinates by using the position of reference stars in the composite images. This package employs the method of the intersection of planes to determine the radiant and reconstruct the trajectory in the atmosphere of meteors recorded from at least two different observing stations (Ceplecha 1987). From the sequential measurements of the video frames and the trajectory length, the velocity of the meteor along its path is obtained. The preatmospheric velocity $V_\infty$ is found by extrapolating the velocities measured at the earliest part of the meteor trajectory. Once this velocity and the atmospheric trajectory are known, the software also calculates the orbital parameters of the corresponding meteoroid.

## 3 RESULTS AND DISCUSSION
The fireball analyzed here (code SPMN081011_194759) with absolute mag. -10.5±0.5 was simultaneously recorded from three of our video meteor observing stations (#1, 2 and 3 in Table 1) on October 8 at 19h47m59.3±0.1 s UTC (Figure 1). This magnitude was determined by comparing the luminosity of the fireball at its maximum brightness with that of nearby stars and the Moon. The bolide was named "Lebrija", because it reached its maximum brightness close to the zenith of this city. After the event took place, a persistent train was clearly visible for several minutes, although this was not imaged because our low-scan high-sensitivity CCD cameras in station #1 could not operate, as was mentioned before. This fireball was similar in brightness to the mag. -11 PN39043 October Draconid fireball, which was imaged by the Prairie Network on Oct. 10, 1965 (Sekanina 1985). However, no emission spectrum was obtained for that event.

### 3.1 Atmospheric trajectory, radiant and orbit
The main parameters characterizing the atmospheric trajectory of the Lebrija fireball can be found in Table 2. The luminous phase began at a height $H_b$ of about 104.7 km above the ground level and terminated at $H_e$=77.0 km. The maximum brightness was observed when the meteoroid was at a height $H_{max}$= 94.5 km. No deceleration was observed between the initial height and ~100 km. Taking this into account, the preatmospheric velocity calculated from the velocities measured at the beginning of the meteor trail was $V_\infty$=23.3 ±0.3 km s$^{-1}$ and the apparent radiant was placed at $\alpha$=269.7±0.2 º and $\delta$=55.1±0.1 º, with the geocentric radiant located at $\alpha_g$=264.1±0.2 º and $\delta_g$=54.6±0.1 º. The orbital parameters were calculated using our AMALTHEA software and the orbit is plotted on Figure 1d. The radiant and orbital parameters are given in Table 3.

### 3.2 Light curve and meteoroid mass
The light curve of the Lebrija fireball, which was obtained from the photometric analysis of the video frames, is shown in Figure 2. For the calculation of this curve we took into account that the detectors at Sevilla and Arenosillo were saturated during the brightest phase of the fireball, but not during its initial and final phases. Just one of the cameras operating at the mobile meteor station (Cerro Negro) did not experience saturation. Its iris, which can be manually regulated, was partially closed on purpose to prevent saturation caused by almost direct moonlight on the CCD sensor. As a result of this, however, this device could not record the fainter phases of the bolide during the beginning and the end of the event. Thus, our light curve was obtained by combining the non-saturated light curve obtained by this device with the averaged (non-saturated) trailing and leading portions of the light curves obtained by the cameras at Sevilla and Arenosillo. This is very smooth compared to the light curve of the mag. -11 PN39043 Prairie Network fireball (Sekanina 1985), with its maximum in the first half of the atmospheric trajectory. Fireballs produced by compact meteoroids reach their maximum brightness later in their atmospheric trajectory (Murray et al. 1999; Campbell et al. 2000) while this takes place earlier for those produced by dustball meteoroids (Hawkes & Jones 1975). This behaviour can be characterized by means of the so-called F parameter (Flemming et al. 1993; Brosch et al. 2004), which is a non-dimensional quantity that is related to the position of the maximum along the light curve. In this case a value of F=0.47 was obtained, which is consistent with the values calculated by other authors for other October Draconid meteors (Koten et al. 2007) and confirms the fragile nature of meteoroids produced by comet 21P/Giacobini-Zinner.

The evolution of the brightness and velocity of the fireball with height has also been used to infer the initial mass of the meteoroid (Novikov et al. 1998; Konovalova 2003). The preatmospheric photometric mass $m_p$ of the fireball was estimated as the total mass lost due to the ablation process between the beginning of the luminous phase and the terminal point of the atmospheric trajectory. This was computed from the equation

$$m_p = 2 \int_{t_b}^{t_e} I_p /(\tau v^2) dt \qquad (1)$$

Here the luminous efficiency is given by $\log(\tau) = -12.50 + 0.17 \log(v)$ for $17 \leq v \leq 27$ km s$^{-1}$ (Ceplecha & McCrosky 1976). In this way, the preatmospheric mass of the particle was estimated to be of about 5.4±0.6 kg. Due to their high porosity (of about 90%), October Draconid meteoroids are known to have densities of about 0.3 g cm$^{-3}$ (Borovička et al. 2007; Babadzhanov & Kokhirova 2009) which would yield a diameter for the meteoroid of about 32 cm.

As was previously mentioned, the mag. -10.5 October Draconid fireball experienced an abrupt increase in luminosity almost from the beginning of its trajectory and exhibited a maximal brightness in the first half of the atmospheric path. This behaviour is consistent with the mechanism of quasicontinuous fragmentation (QFC) described in (Novikov et al. 1984, 1998). The atmospheric density at the height of termination of fragmentation of the meteoroid is lower than the atmospheric density at the height of disappearance of fragments which were detached at the beginning height. A light curve similar to that observed for the mag. -10.5 October Draconid fireball is produced. Such a strong and long flare can be produced by the quasicontinuous fragmentation of meteoroids with a mass higher than $10^{-2}$ - $10^{-3}$ g. This was concluded from the analysis of a very long (1200 m) wake associated to the extraordinarily bright double-station sporadic fireball MK-39 (absolute mag. -12), which was recorded on August 8, 1964 by the method of instantaneous exposure (Babadzhanov 1985). Over most of the trajectory, the shutter breaks were invisible, making this portion look entirely continuous - a phenomenon referred to in meteor physics as blending. The photometric behaviour of this fireball was very similar to that of the Lebrija October Draconid, and also other parameters such as the initial mass, velocity and zenith distance to the radiant.

### 3.3 Tensile strength.

As was previously mentioned, the mag. -10.5 fireball exhibited a strong flare along its trajectory, at a height of about 94.5 km above the ground level. These events are produced by the fragmentation of meteoroids when these particles penetrate towards denser atmospheric regions. With this information the aerodynamic strength at which the particle suffered this break-up can be obtained. The tensile strength S can be inferred from the equation (Bronshten 1981)

$$S = \rho_{atm} \cdot v^2 \qquad (2)$$

where v is the velocity of the meteoroid at the disruption point and $\rho_{atm}$ the atmospheric density at the height where this fracture takes place. This density can be calculated by employing the US standard atmosphere model (U.S. Standard Atmosphere 1976). We infer that the meteoroid exhibited the first flare under a dynamic pressure of $1.9 \pm 0.1 \times 10^2$ dyn cm$^{-2}$. This value, which confirms the fragility of meteoroids belonging to this swarm, is similar to those previously found by other authors for October Draconid meteoroids (Trigo-Rodríguez & Llorca 2006, 2007). This result can be used to estimate the bulk density of the meteoroid by using the graphical fit of bulk density versus compressive strength shown in Figure 1 in (ReVelle 2002) and a value of about 0.1 g cm$^{-3}$ is obtained. As a comparison, the PN39043 fireball exhibited three main flares related to sudden breakup events under pressures of $1.4 \times 10^4$ dyn cm$^{-2}$, $8.6 \times 10^4$ dyn cm$^{-2}$ and $19 \times 10^4$ dyn cm$^{-2}$ (Sekanina 1985). Hence, the bulk density of the PN39043 October Draconid ranged between 0.25 g cm$^{-3}$ and 0.6 g cm$^{-3}$. We conclude that the Lebrija meteoroid had an extraordinarily small bulk density in comparison to the bulk density of the PN39043 meteoroid. Also, a value of 0.1 g cm$^{-3}$ would increase the estimated size of the Lebrija meteoroid from ~32 cm to ~47 cm.

### 3.4 Emission spectrum

The emission spectrum of the Lebrija bolide was obtained by three spectral video cameras. Two of these were located in station #1 and the third one operated from station #3. The first and sec-ond orders are present. However, because of the high brightness of the fireball, the signal corresponding to the first order is saturated during a part of the atmospheric trajectory.

The spectrum has been processed with the CHIMET software (Madiedo et al. 2011b), which follows the analysis procedure described in Trigo-Rodríguez et al. (2003). The video frames containing the emission spectrum were dark-frame substracted and flat-fielded. The signal was calibrated in wavelengths by identifying typical lines appearing in meteor spectra (Ca, Fe, Mg and Na multiplets) and corrected by taking into account the efficiency of the recording instrument.

The most prominent emission lines correspond to Mg I-2 (516.7 nm), Na I-1 (588.9 nm) and several Fe-I multiplets. The contribution from atmospheric $N_2$ is also seen. A differentiated evolution of the intensity of these lines was found, a phenomenon known as differential ablation. The meteoroid is progressively heated, and depending on the volatility of the minerals available at each point of the ablation, some elements are more quickly incorporated to the meteor column than others. The most characteristic feature of this spectrum is the fact the sodium emission starts earlier than that of other elements and is dominant during the first part of the atmospheric trajectory of the fireball. This can be seen in Figure 3, where the evolution of the intensity (in arbitrary units) of the emission lines of Mg and Na is shown as a function of height. These data have been corrected for the spectral response of the imaging device. It must be taken into account that the signal obtained by the spectral camera for first order spectrum is saturated in the region between 96 and 91 km so that this could not be employed to construct these curves within this range. Nevertheless, these were completed with the information inferred from the second order spectrum, which was not saturated due to its lower intensity. Figure 3 also reveals that when the meteoroid penetrates the atmosphere below 98 km, the Mg emission becomes dominant, with the Na line disappearing earlier. This behaviour was previously found by Millman (1972) and Borovička et al. (2007) in two October Draconid meteors, and was explained on the basis of a differential ablation of the grains in the meteoroid (McNeil et al. 1998; Borovicka, 2007). That behaviour for Na was also found in Leonid meteors (Borovicka et al. 1999).

We envision a quite different scenario to explain the Na preferential ablation. Following the Stardust studies of comet 81P/Wild 2 we know that cometary materials are compacted by a fine grained matrix in which small chondrules or inclusions are embedded, not much different than for carbonaceous chondrites (Brownlee et al., 2006). In short period comets like 21P, with frequent perihelion approaches, fine grained materials are preferentially altered by water. At the same time, water mobilizes some elements in the matrix of CM carbonaceous chondrites that can be considered a proxy of cometary materials (Trigo-Rodríguez et al., 2006). Water mobilization of light elements ends with water absorption and formation of precipitates in the empty spaces, and phyllosilicates (Rubin et al., 2007). When a fragile meteoroid penetrates into the atmosphere of Earth it experiences subsequent compaction of the most fragile phases: mostly the fine-grained matrix. Then, compaction could lead to significant volatile outgassing, in which Na, due to its ability to produce light, could trace better this process. Seen from a distance of hundreds of kilometers, the full process will be hidden, and only noticeable by a preferential ablation of Na in the emission spectrum front to other more abundant chondritic elements.

On the other hand, the relative abundances in the radiating plasma of this fireball have been determined at two different instants. These correspond to $t_0+0.12$ s and $t_0+0.62$ s, with $t_0$ being the time of apparition of the bolide. The first of these instants corresponds to early part of the recorded trail, before the fireball reaches its maximum brightness. At the second instant, however, the bolide is located at the final part of its trajectory. These spectra can be seen in Figures 4 and 5 respectively, where the raw and calibrated signals have been included. Due to the low geocentric velocity of the meteoroid, the second (high-temperature) component in the spectrum is almost non-existent as previously described (Borovicka 1994; Trigo-Rodríguez et al. 2003). So, lines identified in Figures 4c and 5c belong to the main component. In order to get the physical parameters in the meteor column we fitted the intensity of the Fe I lines by using the different Fe multiplets identified in the spectrum. This provided the values of the averaged temperature (T) and the Fe I column density (N). Once these values are fixed, we modified the abundances of the other elements identified in the spectrum to get an optimal fit. The typical uncertainties in the determination of these abundances, which obviously depend on the quality of this fit, are in general of about 15% for elements exhibiting bright emission lines (Fe, Na, Mg and Ca) and of about 34% for those with weaker lines. By assuming Fe/Si=1.16 (Anders & Grevesse 1989), we have obtained for $t_0+0.12$ s a value of the averaged temperature of the meteor plasma T=4,000±200 K, with an electron density in the column N=$10^{14}$ cm$^{-2}$. Fe lines were used as a reference as in our previous work (Trigo-Rodríguez et al. 2003, 2004), but we refer here the computed abundance ratios to Si thanks to usually discernible Si I line at 637 nm. Then, the computed abundances for this frame are: Mg/Si=0.61±0.01, Fe/Si=0.65±0.01, Na/Si=0.05±0.01, Ca/Si=0.01±0.01, Cr/Si=0.008±0.001 and Ni/Si=0.09±0.01. On the other hand, for $t_0+0.62$ s we have obtained T=4,400±200 K and N=5·$10^{14}$ cm$^{-2}$, with the following abundance ratios: Mg/Si=0.93±0.01, Fe/Si=0.90±0.01, Na/Si=0.02±0.01, Ca/Si=0.02±0.01, Cr/Si=0.008±0.001, Ni/Si=0.08±0.01 and Mn/Si=0.005±0.001. While the Na is significantly depleted in the ending part of the meteor column, the Fe and Mg contributions are much higher. We suspect that the sustained temperature along the trail is producing an increasing ablation of silicates that are then contributing to increase Mg and Fe into the vapour phase. In general, these relative abundances are consistent with chondritic values and the Na/Mg ratio is in very good agreement with the values obtained by other authors for October Draconid meteoroids (Millman 1972; Borovička 2007). This result is interesting, since a possible application of comparing spectra of bright bolides belonging to different dust trails is the possibility of studying space weathering effects. Na is a volatile element that can be used to trace the effect of meteoroid irradiation in interplanetary space, particularly for orbits exhibiting close perihelion approaches (Trigo-Rodríguez and Llorca 2007b). From the chondritic abundance inferred from Lebrija spectrum, there is no evidence of a bulk Na depletion. Being a member of the 1900 or 1907 dust trails (Trigo-Rodríguez et al. 2012), this meteoroid probably spent a little more than one century in interplanetary space, following an orbit with a very moderate perihelion distance of q=0.9967 (Table 3). It seems that cm-sized 21P meteoroids do not suffer bulk elemental depletions on such timescales. In an astrobiological context, it offers the possibility to preserve pristine cometary materials in the interior of cm-sized meteoroids, also due to the expected thermal inertia of fractal aggregates. Large cometary meteoroids with low entry velocity can fragment in the atmosphere, and their fragments can escape the thermal wave (Trigo-Rodríguez et al. 2012), so they that could be the source under favourable geometric conditions of the recently found ultracarbonaceous micrometeorites (Duprat et al. 2010).

As previously mentioned, five additional October Draconid fireballs (Figure 6) were also recorded by our spectral cameras, although these were single-station events and so no orbits or at-mospheric trajectories were derived from them. These bolides were also imaged during the peak of the 2011 outburst and their brightness ranged from mag. -4.0 to mag. -4.5 (Table 4). Their light curves (Figure 7) indicate that fireballs SPMN081011SA, SPMN081011SC and SPMN081011SE suffered a quasicontinuous fragmentation. However, events SPMN081011SB and SPMN081011SD exhibited flares associated with a sudden disruption. This disruption took place at the end of the atmospheric trajectory for fireball SPMN081011SD. Because of their relatively low luminosity (our spectral cameras can obtain spectra for events brighter than mag. -4), their emission spectrum was imaged just during the brightest phase and so the evolution of the intensity of emission lines with height could not be analyzed. These spectra, once calibrated by following the same procedure described above for the Lebrija fireball, are shown in Figures 8 to 12. As can be noticed, the emission of the Na I-1 multiplet is dominant, with the exception of fireball SPMN081011SE, where the intensity of the Mg I-2 emission line predominates. From this we could infer that SPMN081011SE experienced an early release of Na during the beginning of its atmospheric path. However, as the parent body is expected to be inhomogeneous, this could also be explained by a lower fraction of Na in the meteoroid, or by its having a higher Mg-rich mineral content. This behaviour was also found during the brightest phase of the Lebrija bolide (Fig. 3), whose photometric profile (Fig. 2) is also similar to the light curve obtained for fireball SPMN081011SE. Thus, in both events the maximum luminosity is achieved during the first half of the atmospheric trajectory, with F=0.47 for Lebrija and F=0.48 for SPMN081011SE. On the other hand, the relative abundances of the main chemical elements derived for these spectra are given in Table 5. We should remark that these five spectra were obtained for the peak magnitude seen in Figure 6, and then could be representative of different ablation steps. In any case, by comparing them we have realised important differences in the bulk chemistry of these October Draconid meteoroids. This is not surprising as it should be a consequence of the intrinsic component diversity available in the interior of the mm-sized 21P aggregates that were producing these fireballs (note that all of them were in the –4/-4.5 absolute magnitude range). The Stardust mission demonstrated that most 81P/Wild 2 particles were aggregates formed by diverse components: chondrule-like, and refractory inclusions compacted by a fine grained volatile-rich matrix (Brownlee et al., 2006). A similar picture can be drawn from our data. For example, some spectra like e.g. SPMN081011SA, SPMN081011SB, and SPMN081011SE are quite rich in Fe and Mg, while other like e.g. SPMN081011SC and SPMN081011SD were poor in both of them. On the other hand, Na is slightly over chondritic values for SPMN081011SA and SPMN081011SB, while it is close to the CI chondrite values for the rest of spectra. Finally, a higher abundance of Ca in SPMN081011SA, and SPMN081011SD suggests the presence of a very refractory component for these particles.

## **4 CONCLUSIONS**

We have monitored the October Draconid outburst on October 8, 2011 from different meteor observing stations in Spain, by putting a special focus on meteor spectroscopy. This has provided information about the physical and chemical properties of meteoroids produced by comet 21P/Giacobini-Zinner. The main conclusions of this work are as follows.

1) Bright October Draconid bolides appeared when Earth crossed the most dense trails of comet 21P (Trigo-Rodríguez et al., 2012). The result of the SPMN spectroscopic campaign was the recording of six bolides together with their emission spectra. Unfortunately, only one of them was a multi-station event.

2) We have obtained the trajectory, radiant and orbital elements for a mag. -10.5 double-station October Draconid fireball catalogued as SPMN081011_194759 (Lebrija). The mass inferred from the light curve shows that this bolide was produced by a ~32 cm in diameter meteoroid with a mass of about 5.4±0.6 kg. The inferred tensile strength from the measured height and velocity at its disruption point confirms the fragility of meteoroids produced by comet 21P/Giacobini-Zinner.
3) The emission spectrum produced by the Lebrija fireball has provided the relative abundances of the main elements in the meteoroid. These support a chondritic nature for meteoroids of this stream.
4) October Draconid fireballs exhibit clear signs of differential ablation. For example, Lebrija spectrum shows a clear evolution in the intensity of the emission lines, with an early appearance and disappearance of the Na line. This is probably a consequence of the progressive heating of the meteoroid and the availability of different minerals with height.
5) We suggest a possible mechanism to explain the early appearance of Na. When a fragile meteoroid penetrates into the atmosphere of the Earth it experiences subsequent compaction. If some minerals are hydrated and contain bonded water (e.g. in phyllosilicates), then compaction could lead to water (OH) dehydration and Na preferential outgassing much earlier than other elements make appearance in the emission spectrum. If we are correct, we predict that high-resolution spectroscopy could be able to identify a OH band in the earliest parts of some meteor columns. In particular, cm-sized meteoroids could be good places to retain volatiles.
6) There is no evidence of Na depletion from chondritic abundances in the spectrum of Lebrija bolide. Being a young 21P meteoroid, which probably spent a little bit more than one century in interplanetary space with an orbit with moderate perihelion distance of q=0.9967, our results suggest that cm-sized 21P meteoroids are not experiencing bulk elemental depletions in such timescales. This conclusion can have important implications in astrobiology, as cm-sized meteoroids could have delivered significant amounts of volatiles after cometary disruptions in the early times of Earth's evolution.

**ACKNOWLEDGEMENTS**
We acknowledge support from the Spanish Ministry of Science and Innovation (projects AYA2009-13227, AYA2009-14000-C03-01 and AYA2011-26522), Junta de Andalucía (project P09-FQM-4555) and CSIC (grant #201050I043).

**TABLES**

Table 1. Geographical coordinates of the meteor observing stations involved in this work and the recording devices operating from them.

| Station # | Station | Longitude (W) | Latitude (N) | Alt. (m) | Imaging devices |
|---|---|---|---|---|---|
| 1 | Cerro Negro | 6º 19´ 35" | 37º 40´ 19" | 470 | 2 slow-scan CCD spectrographs |
|   |   |   |   |   | 7 CCD video cameras |
|   |   |   |   |   | 5 CCD video spectrograph |
| 2 | Sevilla | 5º 58´ 50" | 37º 20´ 46" | 28 | 9 CCD video cameras |
|   |   |   |   |   | 1 CCD video spectrograph |
| 3 | El Arenosillo | 6º 43´ 58" | 37º 06´ 16" | 40 | 4 CCD video cameras |
|   |   |   |   |   | 4 CCD video spectrograph |
|   |   |   |   |   | 1 all-sky slow-scan CCD camera |
| 4 | Sierra Nevada | 3º 23´ 05" | 37º 03´ 51" | 2896 | 5 CCD video cameras |
|   |   |   |   |   | 4 CCD video spectrographs |

Table 2. Trajectory and radiant data for the SPMN081011_194759 **October Draconid** fireball.

| $H_b$ (km) | $H_{max}$ (km) | $H_e$ (km) | $\alpha_g$ (º) | $\delta_g$ (º) | $V_\infty$ (km s$^{-1}$) | $V_g$ (km s$^{-1}$) | $V_h$ (km s$^{-1}$) |
|---|---|---|---|---|---|---|---|
| 104.7 | 94.5 | 77.0 | 264.1 ±0.2 | 54.6 ±0.1 | 23.3 ±0.5 | 20.6 ±0.5 | 39.2 ±0.5 |

Table 3. Orbital data (J2000) for the SPMN081011_194759 **October Draconid** bolide. Note that its descending node is almost identical that the forecasted timing of the 1907 dust trail encounter.

| a (AU) | e | i (º) | Ω (º) | ω (º) | q (AU) |
|---|---|---|---|---|---|
| 3.7±0.3 | 0.73±0.03 | 31.0±0.4 | 195.1872±10$^{-4}$ | 173.8±0.1 | 0.9967±0.0001 |

Table 4. Single-station fireballs recorded on Oct. 10, 2011 together with their emission spectra.

| Code | Time (UT) | Visual mag. |
|---|---|---|
| SPMN081011SA | 19h18m38.3±0.1s | -4.0±0.5 |
| SPMN081011SB | 19h34m35.5±0.1s | -4.5±0.5 |
| SPMN081011SC | 19h51m57.1±0.1s | -4.0±0.5 |
| SPMN081011SD | 20h18m47.8±0.1s | -4.0±0.5 |
| SPMN081011SE | 20h32m32.6±0.1s | -4.5±0.5 |

Table 5. Relative abundances for the main rock-forming elements of the single-station fireballs listed in Table 4.

| Code | T (K) ±100 | N (cm$^{-3}$) | Mg/Si | Na/Si | Fe/Si | Ca/Si | Cr/Si | Mn/Si |
|---|---|---|---|---|---|---|---|---|
| SPMN081011SA | 4500 | 5·10$^{14}$ | 1.08 | 0.08 | 0.90 | 3.5·10$^{-2}$ | 1·10$^{-2}$ | 5·10$^{-3}$ |
| SPMN081011SB | 4300 | 1·10$^{14}$ | 1.17 | 0.10 | 0.85 | 2.0·10$^{-2}$ | - | - |
| SPMN081011SC | 4000 | 5·10$^{13}$ | 1.06 | 0.07 | 0.62 | 4.0·10$^{-2}$ | - | 1·10$^{-2}$ |
| SPMN081011SD | 4200 | 1·10$^{13}$ | 0.86 | 0.06 | 0.71 | 3.8·10$^{-2}$ | 5·10$^{-3}$ | 5·10$^{-3}$ |
| SPMN081011SE | 4200 | 1·10$^{14}$ | 1.13 | 0.04 | 0.80 | 3.8·10$^{-2}$ | 5·10$^{-3}$ | - |

**FIGURES**

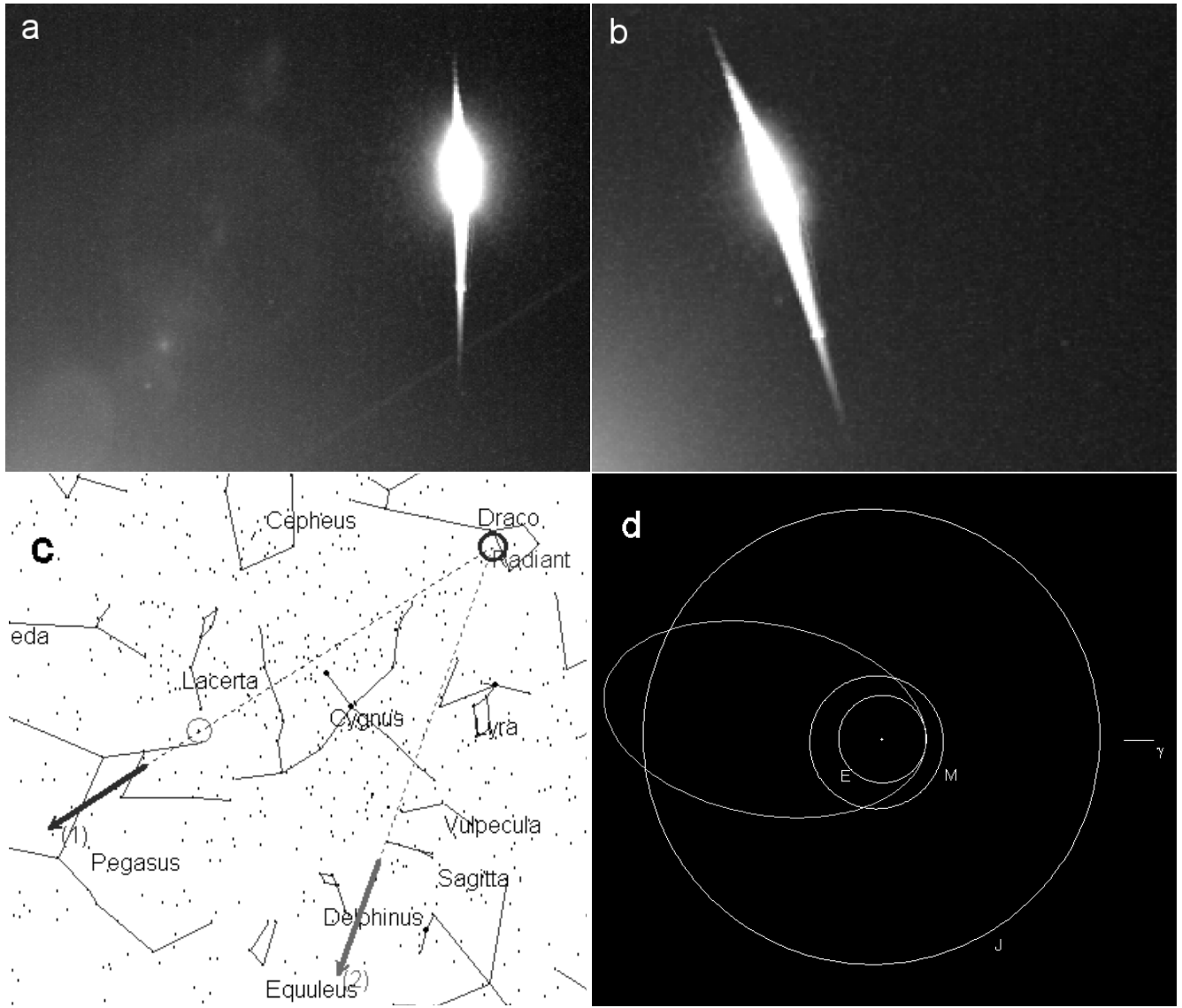

Figure 1. The Lebrija bolide (SPMN081011_194759), with an absolute magnitude of −10.5±0.5, recorded from a) Sevilla and b) Cerro Negro. c) Apparent trajectory of the bolide as seen from Sevilla (1) and Cerro Negro (2). d) Heliocentric orbit of the meteoroid projected on the ecliptic plane.

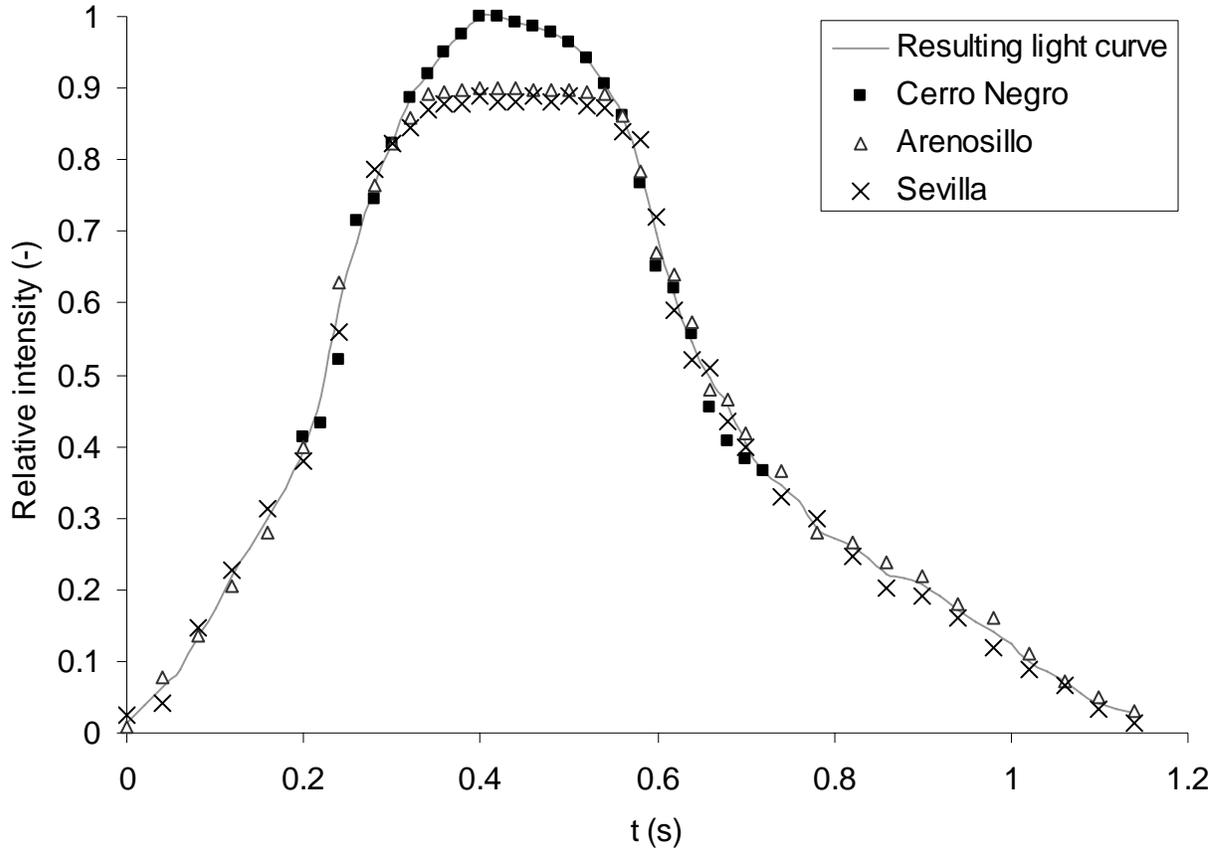

Figure 2. Light curves (relative brightness vs. time) obtained from the devices operating from Sevilla, Arenosillo and Cerro Negro meteor observing stations for the SPMN081011_194759 October Draconid fireball. The continuous line corresponds to the resulting (averaged) light curve by taking into account the saturation of the devices at Sevilla and Arenosillo during the maximum phase of the event.

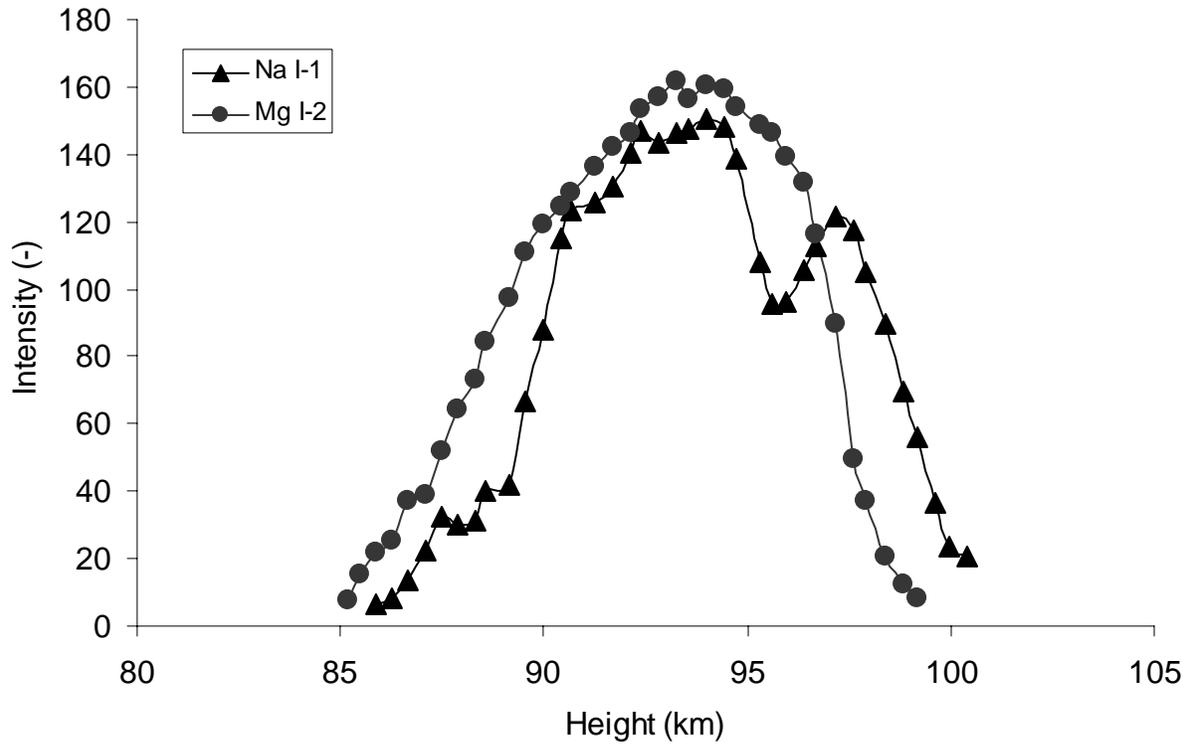

Figure 3. Intensity (in arbitrary units) as a function of height for the emission lines corresponding to multiplets Mg I-2 (516.7 nm) and Na I-1 (588.9 nm) in the spectrum of the Lebrija fireball.

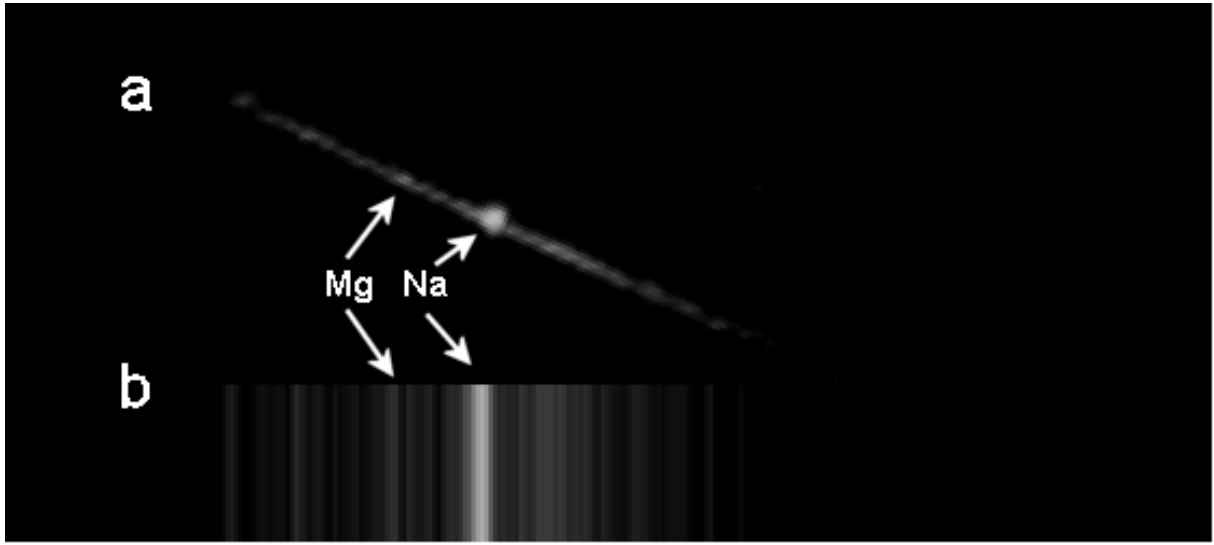
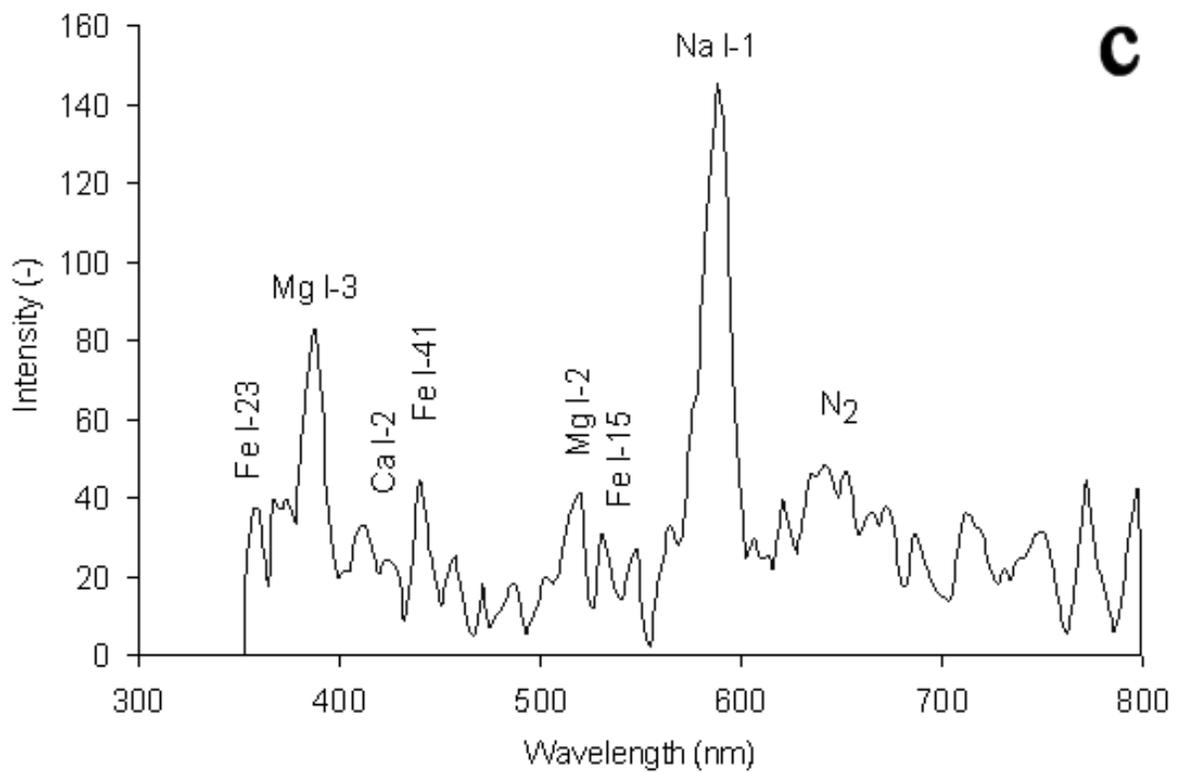

Figure 4. a) Raw and b) calibrated emission spectrum of the Lebrija fireball at time $t=t_0+0.12$ s ($t_0$ is the instant of meteor apparition). c) Main lines identified in the spectrum, together with their multiplet number (intensity is expressed in arbitrary units).

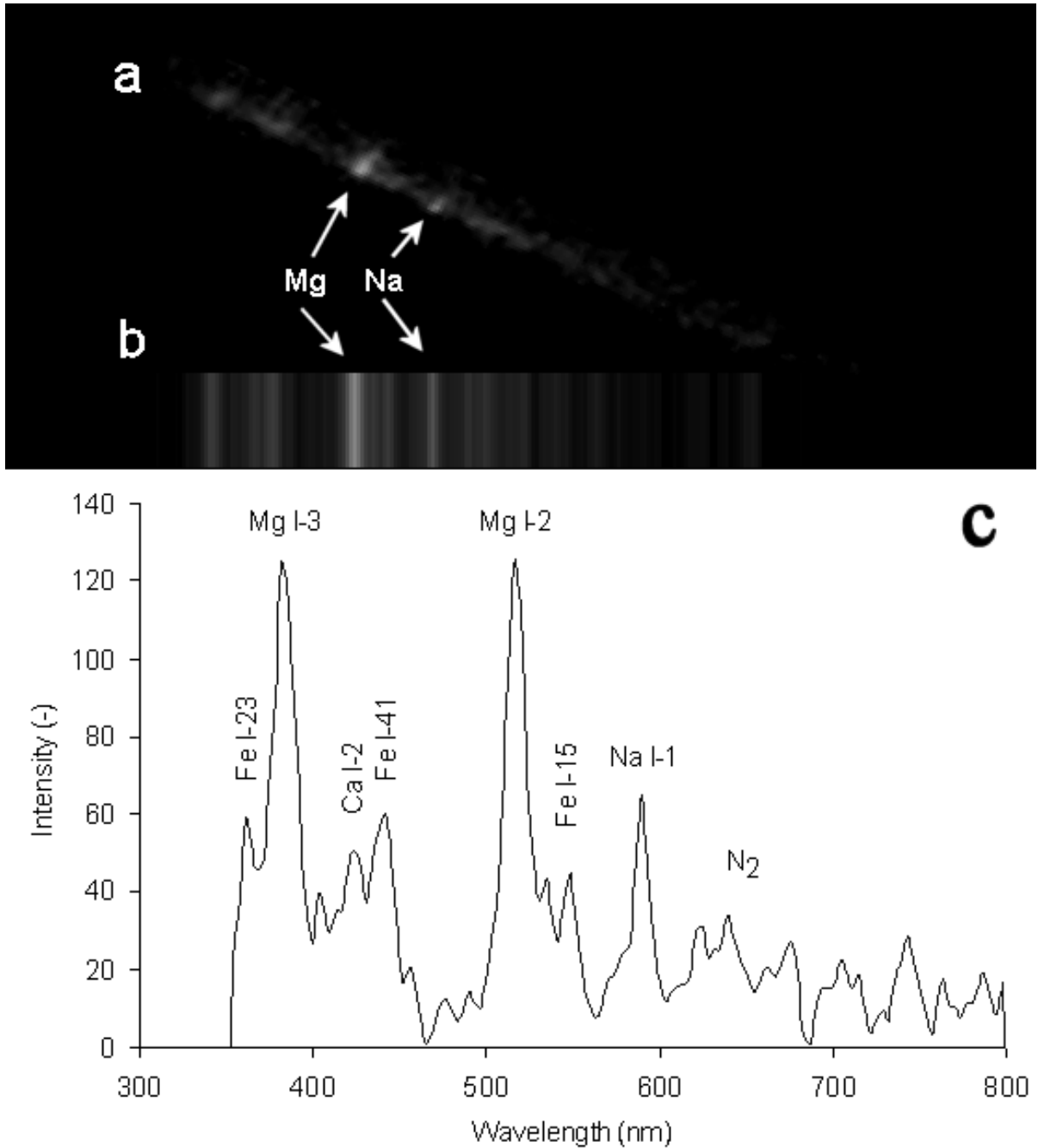

Figure 5. a) Raw and b) calibrated emission spectrum of the Lebrija fireball at time $t=t_0+0.62$ s ($t_0$ is the instant of meteor apparition). c) Main lines identified in the spectrum, together with their multiplet number (intensity is expressed in arbitrary units).

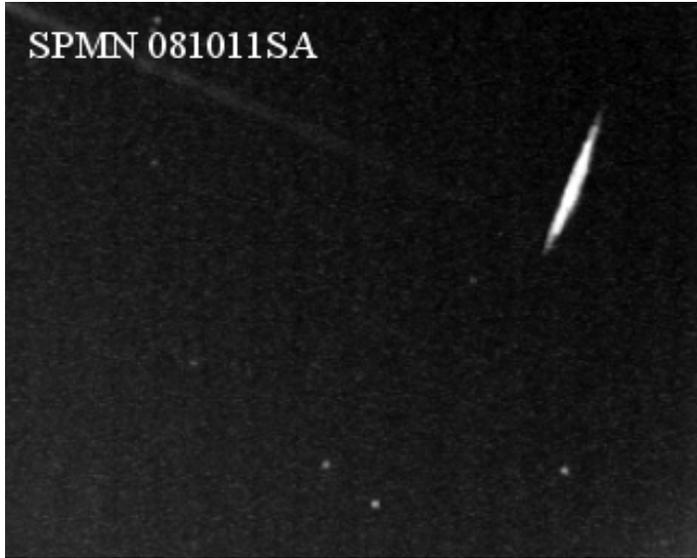
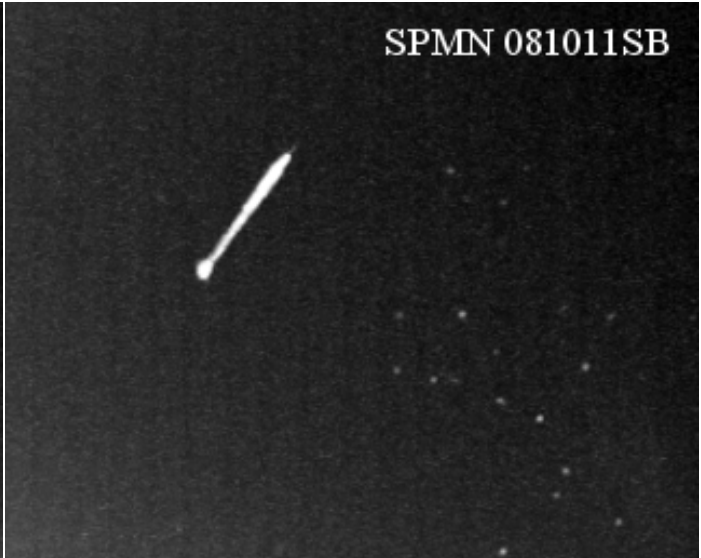
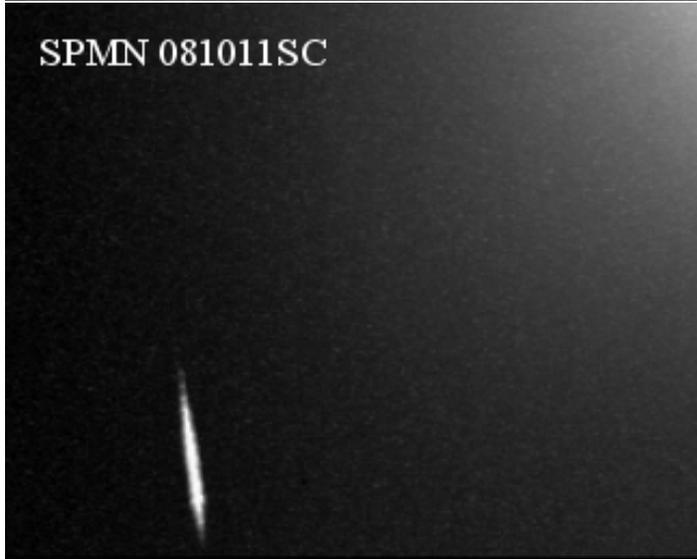
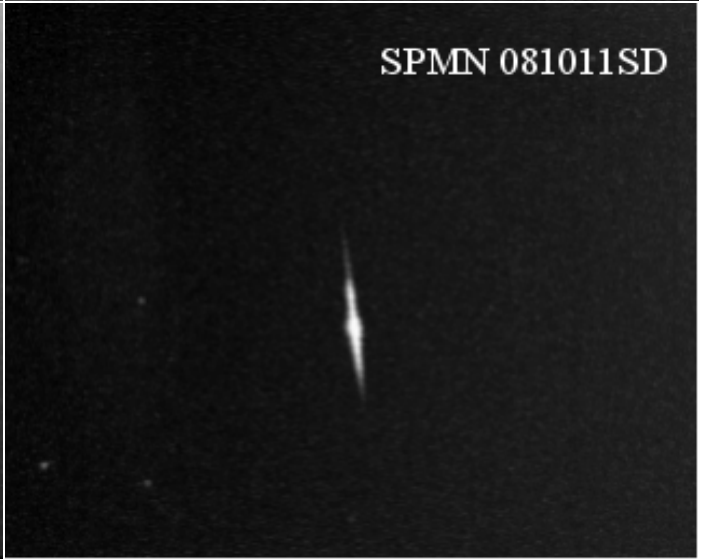
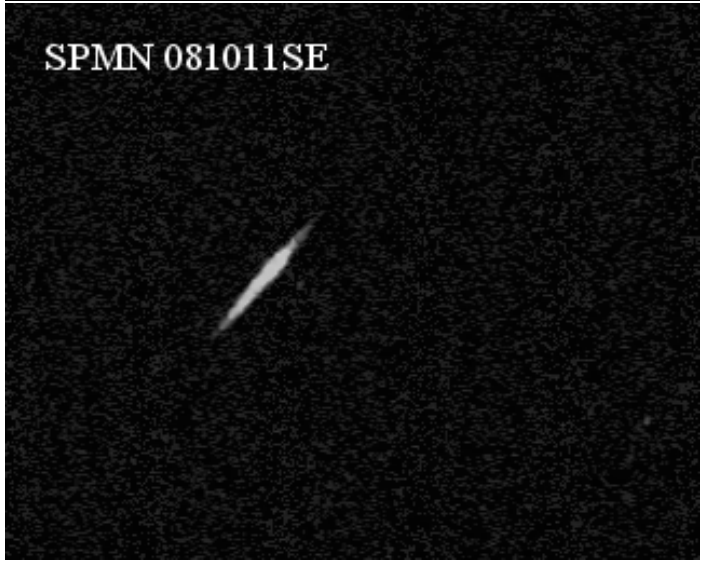

Figure 6. Composed images of the five single-station October Draconid fireballs listed in Table 4.

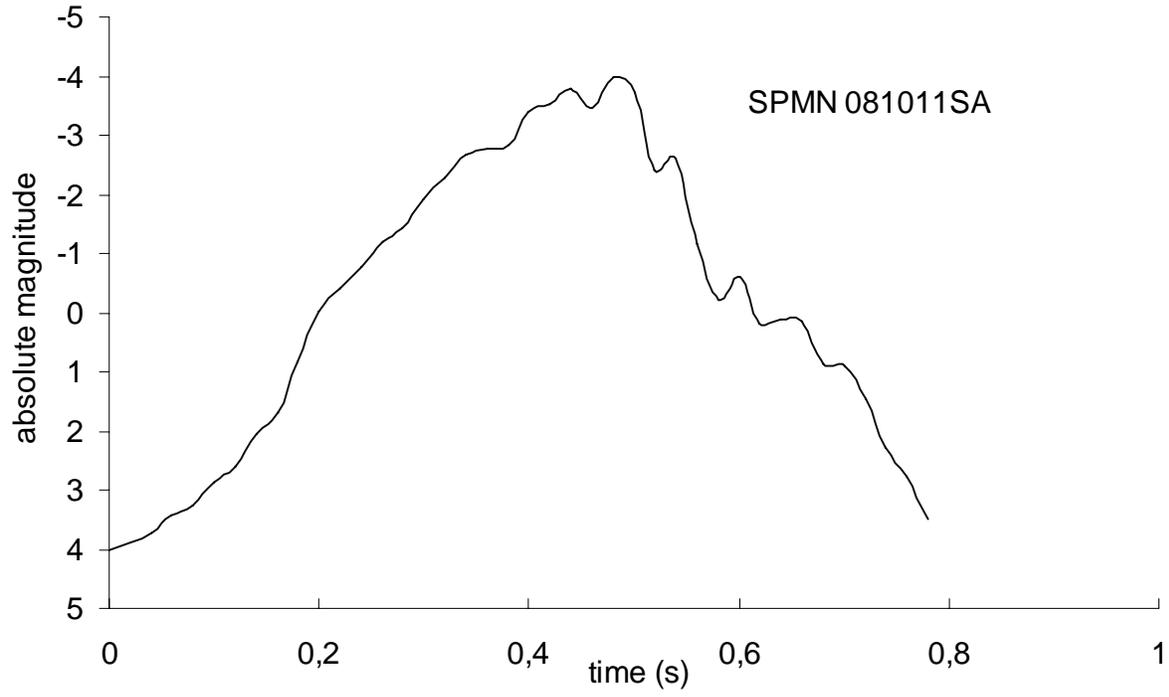
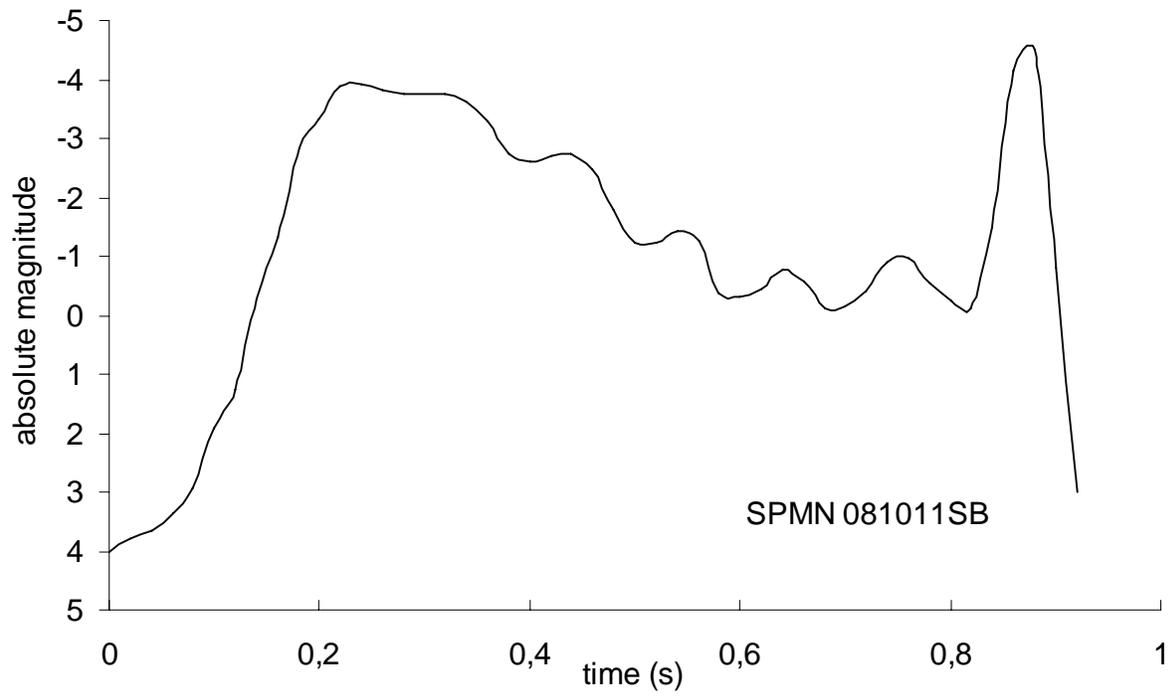

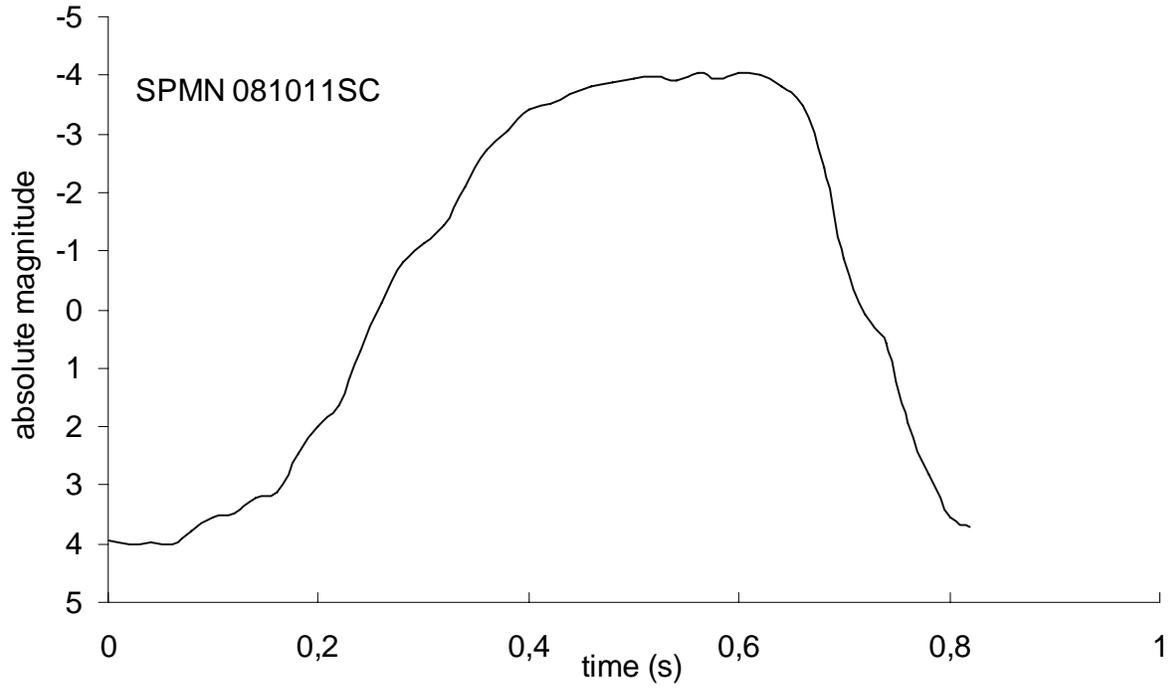
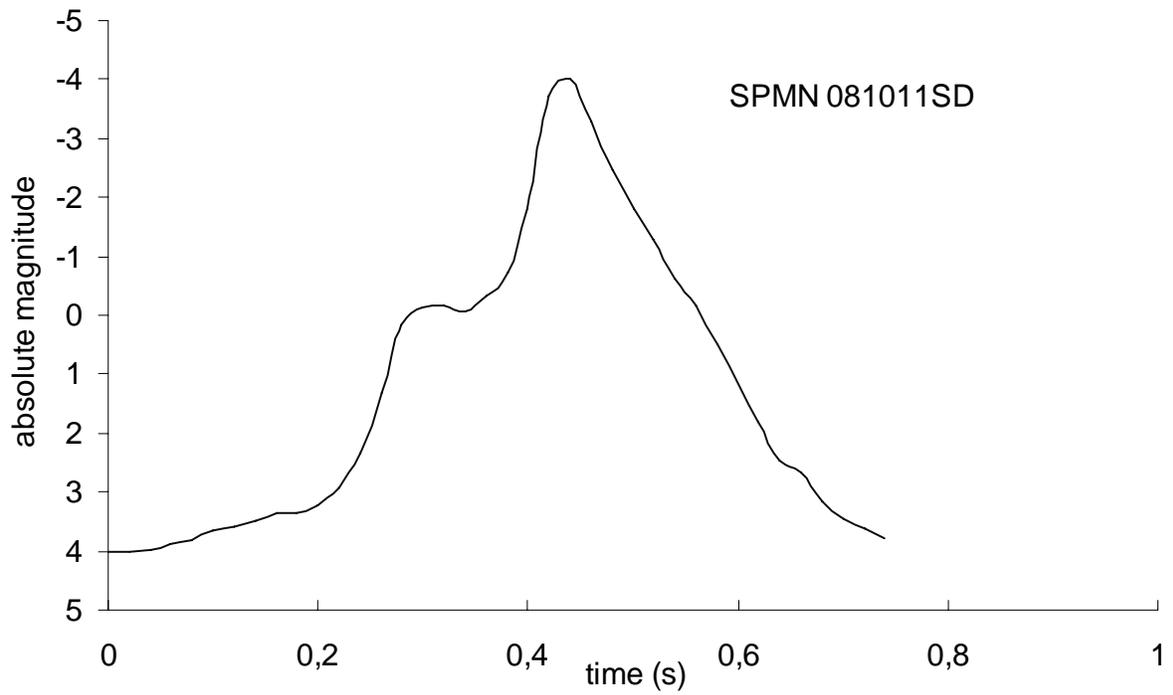

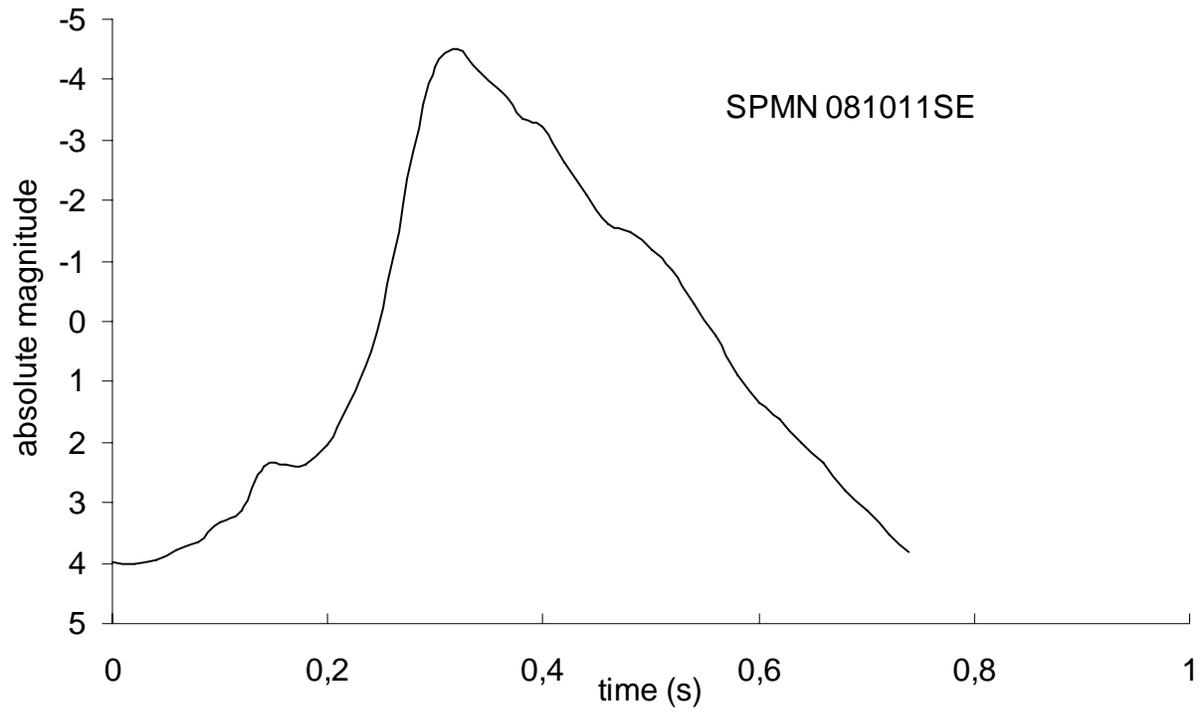
Figure 7. Light curves of the five single-station October Draconid fireballs listed in Table 4.

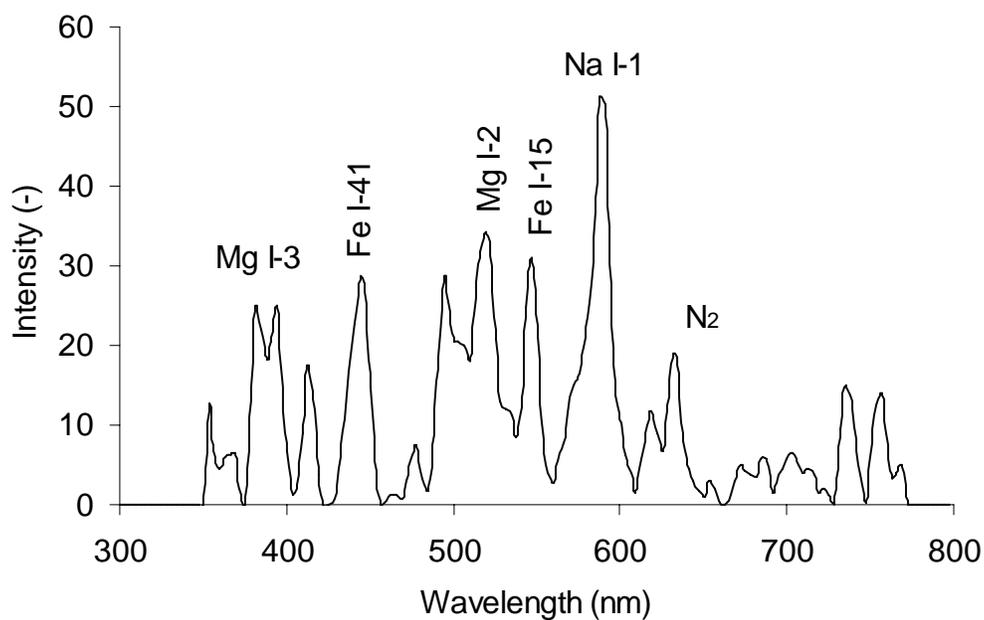

Figure 8. Calibrated emission spectrum recorded for the SPMN081011SA fireball. Intensity is expressed in arbitrary units.

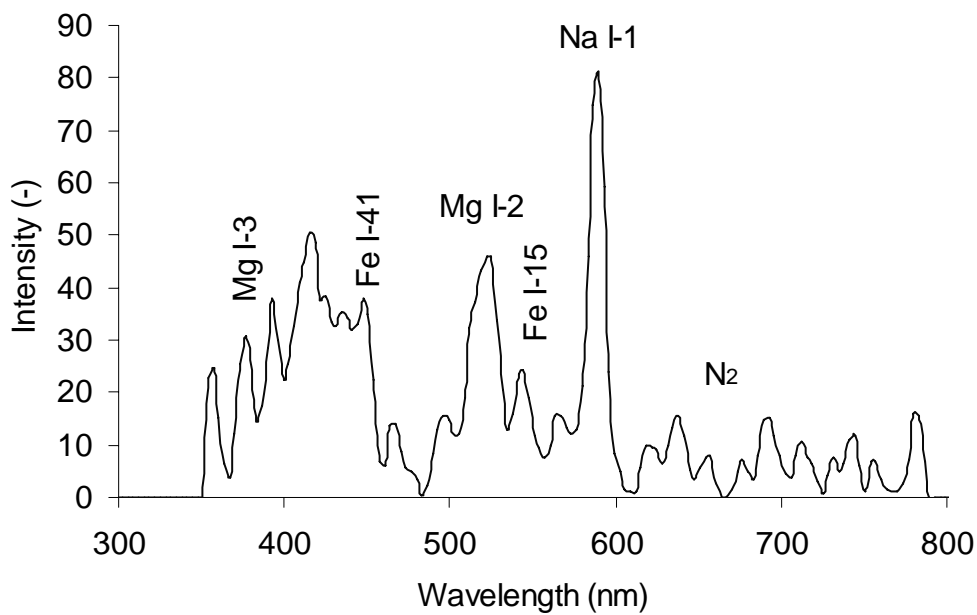

Figure 9. Calibrated emission spectrum recorded for the SPMN081011SB fireball. Intensity is expressed in arbitrary units.

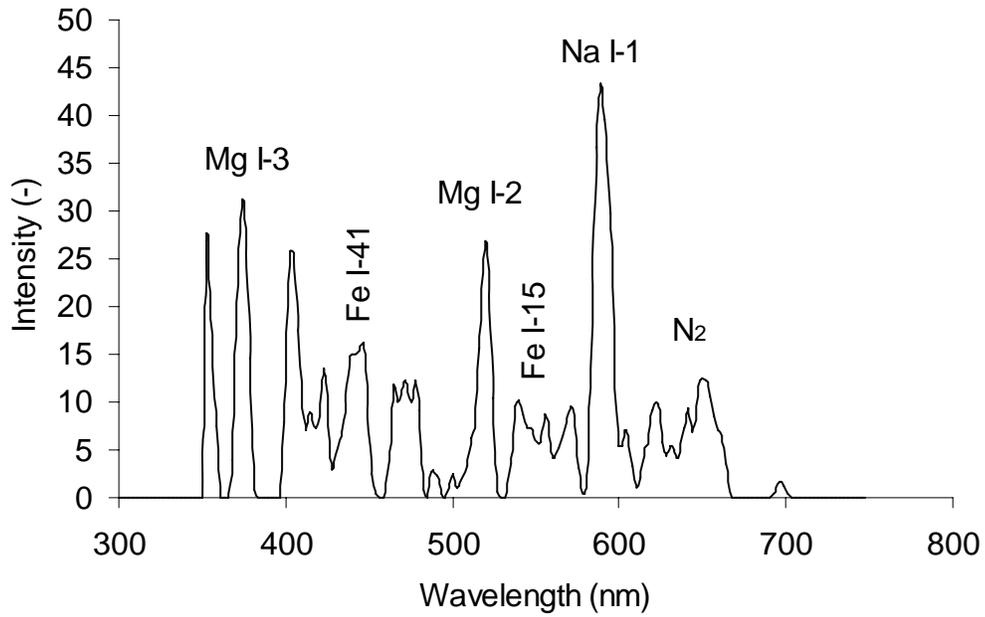

Figure 10. Calibrated emission spectrum recorded for the SPMN081011SC fireball. Intensity is expressed in arbitrary units.

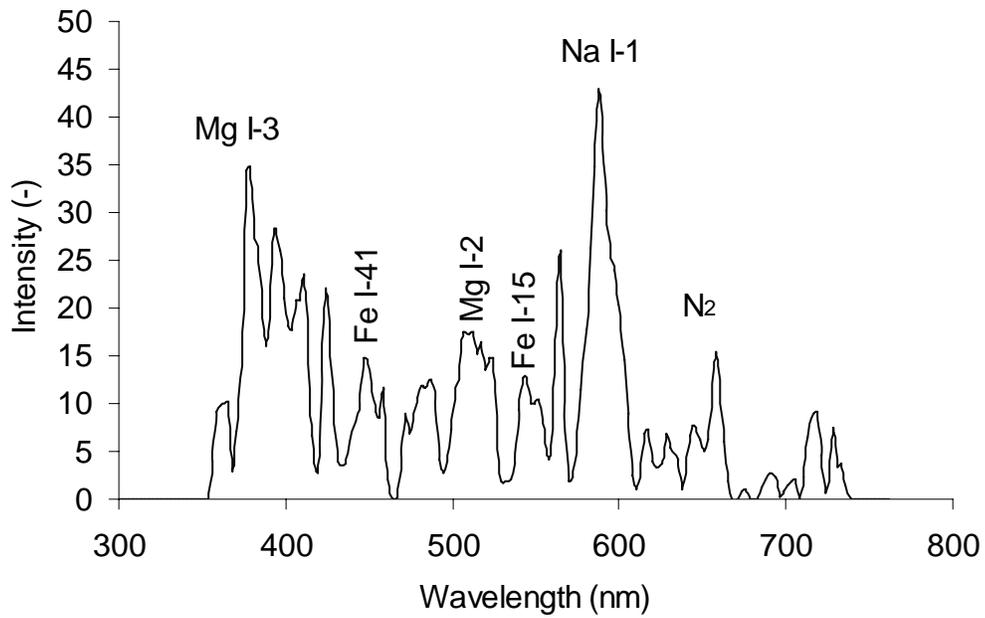

Figure 11. Calibrated emission spectrum recorded for the SPMN081011SD fireball. Intensity is expressed in arbitrary units.

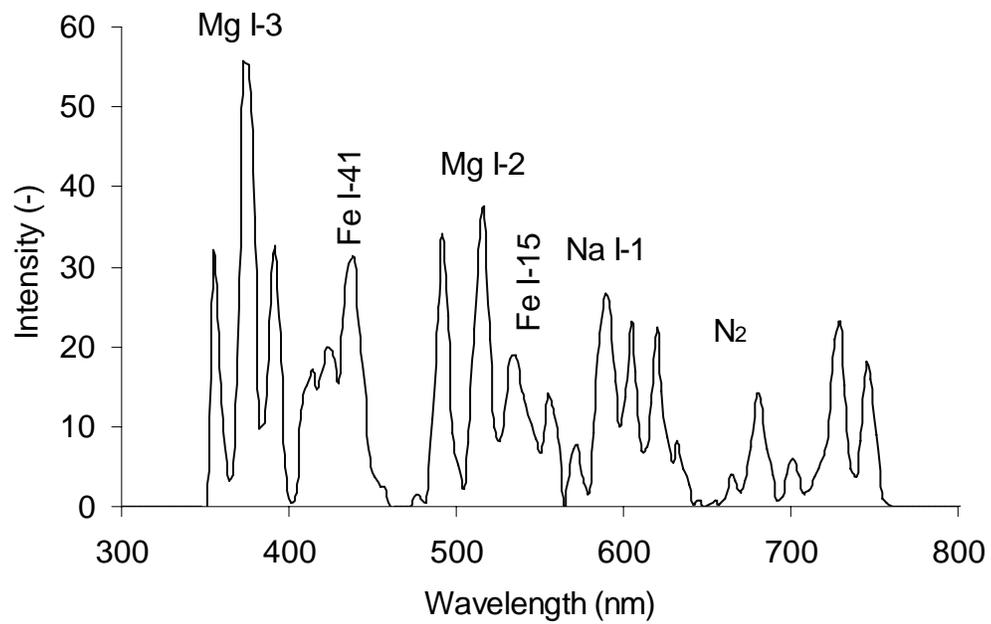

Figure 12. Calibrated emission spectrum recorded for the SPMN081011SE fireball. Intensity is expressed in arbitrary units.